\documentclass[preprint,superscriptaddress]{revtex4}
\usepackage{graphicx}  
\usepackage{dcolumn}   
\usepackage{bm}        
\usepackage{amssymb}   
\usepackage{amsmath}
\usepackage{subfigure}

\newcommand{\unit}[2]{$#1 \, \text{#2}$}

\begin{document}

\title{Direct determination of the transition to localization of light in three dimensions}

\author{T.~Sperling} \affiliation{Fachbereich Physik, University of Konstanz,
Universit\"atsstrasse 10, 78457 Konstanz, Germany}
\author{W.~B{\"u}hrer} \affiliation{Fachbereich Physik, University of Konstanz,
Universit\"atsstrasse 10, 78457 Konstanz, Germany}
\author{C.M.~Aegerter} \affiliation{Physik-Institut, University of Z{\"u}rich, Winterthurerstrasse 190, 8057 Z\"urich, Switzerland}
\author{G.~Maret} \affiliation{Fachbereich Physik, University of Konstanz,
Universit\"atsstrasse 10, 78457 Konstanz, Germany}

\date{\today}

\maketitle

{\bf In the diffusive transport of waves in three dimensional media, there should be a phase transition with increasing disorder to a state where no transport occurs. This transition was first discussed by Anderson in 1958 \cite{anderson58} in the context of the metal insulator transition, but as was realized later it is generic for all waves \cite{anderson85,john84}. However, the quest for the experimental demonstration of "Anderson" or "strong" localization of waves in 3D
has been a challenging task. For electrons \cite{bergmann} and cold atoms \cite{kondov11}, the challenge lies in the possibility of bound states in a disordered potential well. Therefore, electromagnetic and acoustic waves have been the prime candidates for the observation of Anderson localization
\cite{kuga,albada,wolf,drakegenack,wiersma97,scheffold99,fiebig08,stoerzer06,stoerzer06n2,acousticexp,hu08}. The main challenge using light lies in the distinction between effects of
absorption and localization \cite{wiersma97,scheffold99}. Here we present measurements of the time-dependence of the transverse width of the intensity distribution of the transmitted waves, which provides a direct measure of the localization length and is independent of
absorption. From this we find direct evidence for a localization transition in three dimensions and determine the corresponding localization lengths.}

In the diffusive regime ($kl^* \gg 1$) the mean square width
$\sigma^2$ of the transmitted pulse, i.e. the spread of
the photon cloud, is described by a linear increase in time
$\sigma^2 = 4 D t$ \cite{lenke00}. Here, $D$ is the diffusion
coefficient for light, $k$ the wave-vector and $l^*$ the transport mean free path. When considering
interference effects of the diffusive light, Anderson et. al \cite{abrahams} predicted a transition
to localization in three dimensional systems at
high enough turbidity $(kl^*)^{-1}$. The criterion for where this
transition should occur is known as the Ioffe-Regel criterion, namely $kl^* \lesssim 1$ \cite{ioffe60}. At such high turbidities,
light will be localized to regions of a
certain length scale, namely the localization length $\xi$, which
diverges at the transition to localization. This implies that $\sigma^2$ initially increases linearly with time, but saturates at a later time $t_{\text{loc}}$ (localization time) towards a constant value given by
 $\sigma^2 = \xi^2$, where $\xi$ is the localization length.

In this work we present measurements of light propagation in 3D
open, highly scattering TiO$_2$ powders. Given the high turbidity of the samples studied and the large slab
thickness ($L$ varying from 0.6 mm to 1.5 mm) the transmitted
light undergoes typically a few million scattering events in any of the three spatial directions before leaving
the sample. Thus our samples present a true bulk 3D medium for light transport.

\begin{figure}
    \subfigure{\includegraphics[width=0.32\linewidth]{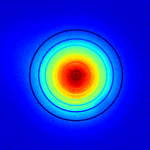}}
    \subfigure{\includegraphics[width=0.32\linewidth]{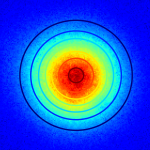}}
    \subfigure{\includegraphics[width=0.32\linewidth]{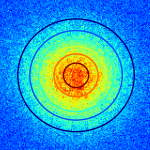}}
    \caption{\label{fig:raw} This figure shows a section of the raw data (with the
    fit displayed via contours) of a R104 sample that features a plateau. From left
    to right the timestamps are \unit{4}{ns}, \unit{6}{ns} and \unit{8}{ns} after
    the initial laser-pulse. From \unit{4}{ns} to \unit{6}{ns} one can see a
    broadening in the profile width, whereas from \unit{6}{ns} to \unit{8}{ns}
    no further increase can be seen. This constant profile width is the signature
    of Anderson localization.}
\end{figure}

The great advantage of determining the time dependence of the
width of the transmission profile lies in the fact that since the
width is obtained at a specified time, absorption effects are present on all paths equally. This means that the width of
the profile at a given time is {\em independent of absorption}. This can
be seen from the general definition of the width in terms of the
spatial dependence of the photon density $T(\rho,t)$, where $\rho$
is a vector in the 2D transmission plane with the origin at the
center of the beam: $\sigma^2(t) = \int\rho^2 T(\boldsymbol{\rho},t)\text{d}^2\boldsymbol{\rho}/\int T(\boldsymbol{\rho},t)\text{d}^2\boldsymbol{\rho}.$
In this definition, an exponential decrease due to absorption enters $T(\rho,t)$ both in the
nominator and in the denominator and thus cancels out.
In the diffusive regime, the profile will be given by a Gaussian:
$T(\rho) \propto \exp(-\frac{\rho^2}{8Dt})$, i.e. with a width
$\sigma^2 = 4Dt$. Hence we fit a 2D Gaussian to the intensity
profile at a given time (see Fig. \ref{fig:raw}, which shows the gated intensity profile at three different time points demonstrating the increase in width with time). This fit yields the
width of the Gaussian in both the x- and y-direction. In localizing samples, the intensity distribution is expected to be exponential, with a characteristic length scale $\xi$. This can be seen in our samples, however at small distances $\rho$, the profile can be well approximated by a Gaussian (see supplementary material). Hence, we fit a Gaussian to all our samples, which gives qualitatively similar fits as an exponential function in the localized case (see supplementary material).

\begin{figure}
    \includegraphics[width=1\linewidth]{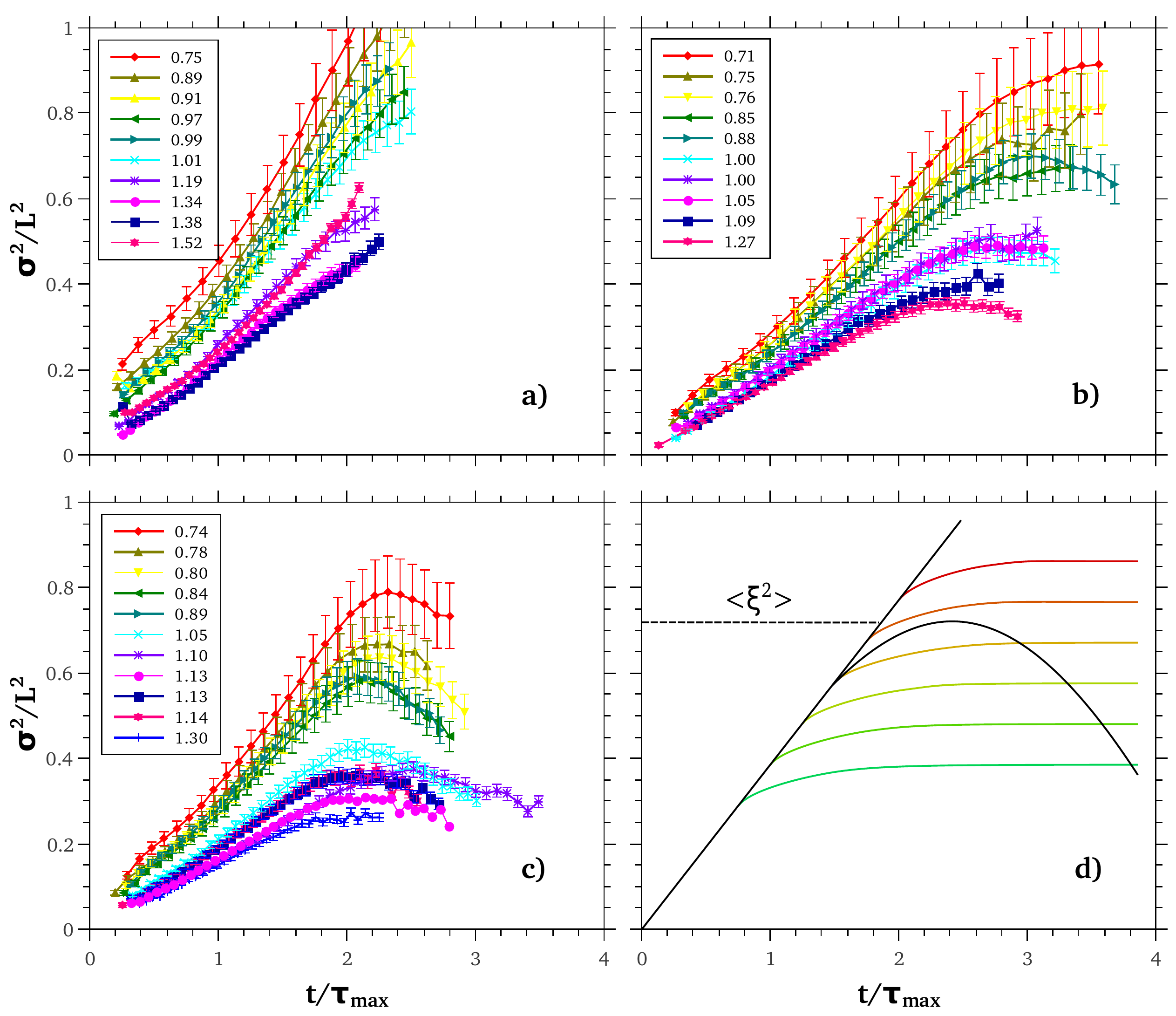}
    \caption{\label{fig:tmax} The mean square width scaled with the sample
    size $\frac{\sigma^2}{L^2}$ is shown for different samples. The time axis
    is scaled with the diffusion time $\tau_\text{max}$ (see supplementary material). In a) Aldrich anatase
    is shown, which behaves diffusively. Samples showing localizing effects
    are b) R104 and c) R700. The legends show the slab thickness $L$ in mm. d) Schematic illustration of the expectation for the time dependence of the width in the presence of statistically distributed localization lengths as discussed in the text. The decreasing population of the different grey lines at late times for larger localization lengths, leads to an overall decrease of the width in particular for sample thicknesses close to the average localization length, because big loops are leaking out of the sample. The different coloured lines correspond to the time dependence of the width with increasing microscopic localization length from small (green) to large (red). }
\end{figure}

The fitted widths are then plotted as a
function of time to yield the results shown in Fig.
\ref{fig:tmax}. In the case of a diffusive sample, Aldrich
anatase, with $kl^*_{AA} = 6.4$, the square of the width
increases linearly over the whole timespan (see Fig.
\ref{fig:tmax} a) as expected. The small deviation from
linearity around the diffusion time $\tau_{max}$ is a result of the
gating of the high rate
intensifier (HRI) \cite{HRI} (see supplementary material). The slope of the increase is in
very good accord with the diffusion coefficient determined from
time dependent transmission experiments \cite{stoerzer06}. Note also that the time dependent width can
exceed the thickness of the sample, which is a consequence of
the fact that we are studying the transmission profile at
specific times.

The width $\sigma^2$ of the transmitted pulse
gives a direct measure of the localization length $\xi$ in the
localizing regime. This is because the 2D transmission profile of
the photon cloud is confined to within a localization length. When considering an effective diffusion coefficient corresponding to the slope of the temporal increase in width, one thus obtains an effective decrease of the diffusion
coefficient with time as $D(t) \propto 1/t$ after a time scale
corresponding to the localization length \cite{berkovits}. In
this picture, for large $L$, one expects a time dependence of the width,
which is linear up to the localization length and then remains
constant as time goes on. Numerical calculations of
self-consistent theory \cite{skipetrov06,cherroret10} give a
different increase at short times as $\sigma^2 \propto t^{1/2}$
and a plateau value of $\sigma^2 = 2L\xi$ for $L \gg \xi$. These predictions can
be directly tested from data of samples with high turbidity, which
show non-classical diffusion in time dependent transmission
measurements. This is shown in Fig. \ref{fig:tmax} b) and c). Taking a
closer look at the short time behavior one can see that $\sigma^2$ increases linearly in time contrary to the
self-consistent theory calculation. This is similar to the
behavior found in acoustic waves \cite{hu08}. However
in contrast to the diffusive sample, a plateau of the width
can be clearly seen. This is in good accord with the theoretical
prediction and a direct sign of Anderson localization. This
plateau can also be seen directly from the transmission profiles
shown in Fig. \ref{fig:raw}, where the normalized intensity
profile is shown for three different time points. At late times,
the width does no longer increases indicating a localization of
light.


The data shown in Fig. \ref{fig:tmax} also show results for
samples of different thickness. These samples of different
thickness are made from the same particles but may vary slightly
in terms of filling fraction. However as checked by coherent
backscattering, samples made up from the same particles have very
comparable turbidity (see supplementary material). If the thickness $L$ of the sample becomes comparable to the localization length, a decrease of the width of the photon distribution with time can be observed. This surprising fact can be understood in a statistical picture of localization, where a range of localization lengths exists in the sample corresponding to different sizes of closed loops of photon transport. In finite slabs, larger localized loops will be cut off by the surfaces leading to a lower population of such localized states at longer times. Thus on average, the observed width will correspond to increasingly shorter localization lengths and thus a decrease of $\sigma^2$ with time can be observed. This is schematically illustrated in Fig.~\ref{fig:tmax} d). Such a peak in the width of the intensity distribution has also been seen in calculations of self-consistent theory, albeit in thicker samples \cite{cherroret10}.
When the thickness decreases even more, such that it is shorter than the localization length, the plateau in the width is lost altogether and $\sigma^2$
increases over the whole time window. In fact, the behavior then corresponds to that predicted for the mobility edge \cite{berkovits}, where a sub-linear increase of $\sigma^2 \propto t^{2/3}$ is predicted. At the transition one observes a kink in $\sigma^2$ and the ratio of the initial slope to that at the kink corresponds to the sub-diffusive exponent $a$. In addition, this thickness
dependence can be used as an alternative determination of the
localization length.

The evaluation of the plateaus of the localizing samples for different thicknesses, yields a localization length independent of $L$.
In case the time dependence showed a maximum rather
than a plateau, the maximum value was used. Thus we identify $\sigma_\infty^2 = \xi^2$ and obtain
$\xi_\text{R104} = 717(6) \mu\text{m}$ for R104, $\xi_\text{R902}
= 717(9)\mu\text{m}$ for R902 and $\xi_\text{R700} =
670(9)\mu\text{m}$ for R700. These are mean values for all
thicknesses investigated.

As expected, sample R700 has the smallest
localization length $\xi$, as has already been concluded by time
of flight experiments \cite{stoerzer06} and corresponds to the
lowest value of $kl^*_{R700} = 2.7$ in this sample. In terms of localization,
R104 and R902 are very similar, which again is in good accord with
the fact that their turbidities are very similar, $kl^*_{R104} = 3.7$ and $kl^*_{R902} = 3.4$ respectively, even though their
other sample properties are rather different. As stated above,
this determination of $\xi$ is in good accord with that from the
thickness dependence of the occurrence of a plateau. As seen in
Fig. \ref{fig:tmax}, R104 with a thickness of $L = 0.71\text{mm}$
behaves sub-diffusively, but the sample with $L = 0.75\text{mm}$
shows a plateau, indicating a localization length of $\xi =
0.73(2)\text{mm}$. The same transitional behavior can be seen for R902
between $0.7\text{mm}<L<0.8\text{mm}$ as well.

\begin{figure}
\includegraphics[width=\linewidth]{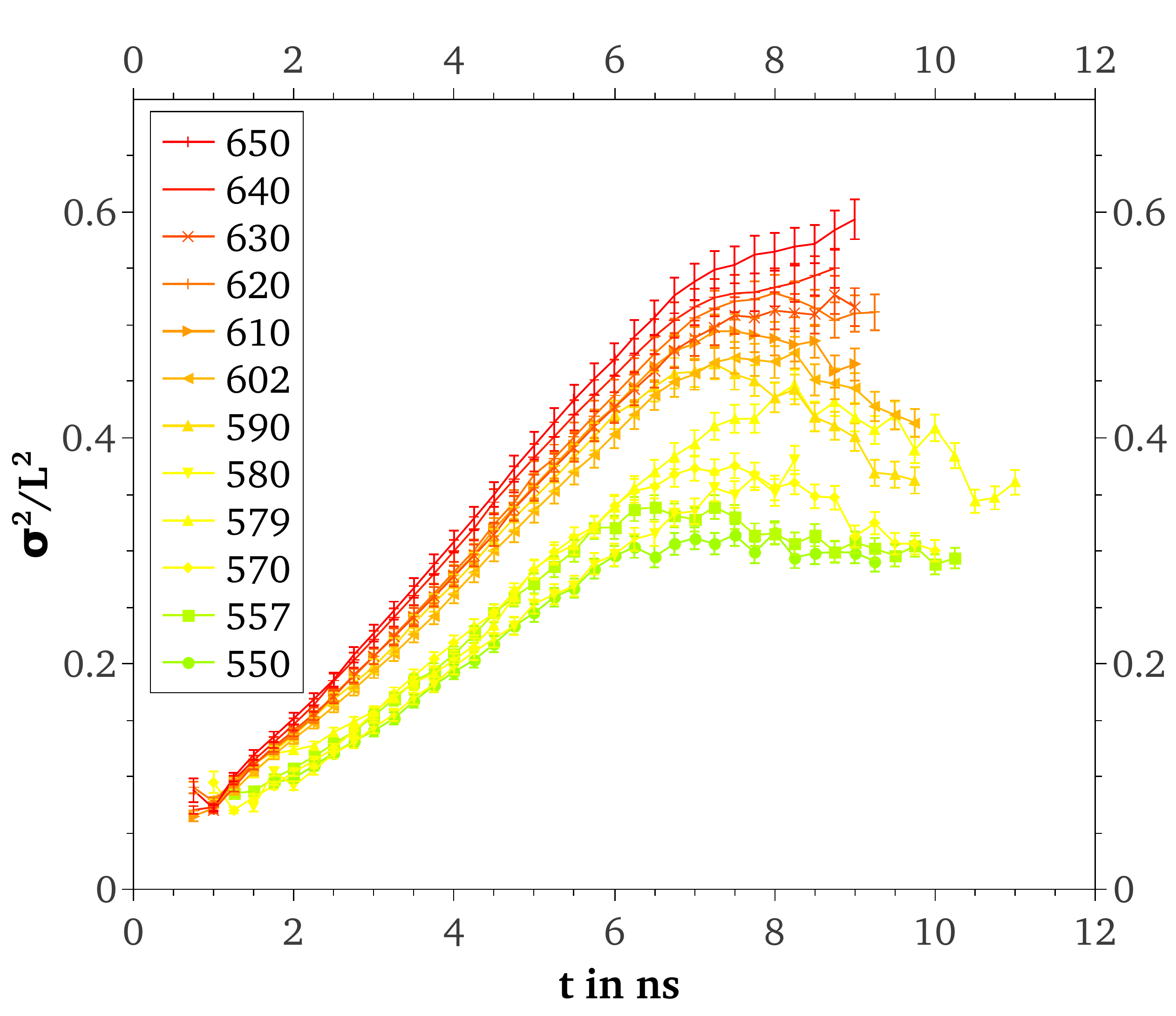}
\caption{\label{fig:spectral} The spectral measurement
of a R700 sample ranging from \unit{550}{nm} -- \unit{650}{nm}, corresponding to $kl*$ values between 2.1 and 3.6,
is shown. With decreasing wavelength $\lambda$ the turbidity
$kl^*$ increases, as well as localizing effects are getting
stronger. This can be seen via the lower mean square width
$\sigma^2_\infty$ of the plateaus. For the wavelengths above
\unit{640}{nm} one can observe a breakdown of localization
to a sub-diffusive behavior. The legend shows the wavelength of light in nm.}
\end{figure}

So far, we have shown that for different samples showing a range
of $kl^*$ close to unity a qualitative change in the transport
properties occurs which is consistent with the transition to Anderson localization. In order to show that these are not sample intrinsic
properties, we now study one and the same sample at different
incoming wavelengths. The turbidity depends quite strongly
on the wavelength $\lambda$ of light, which we tuned from
\unit{550}{nm} to \unit{650}{nm}. For these wavelengths, we have
determined that the turbidity changes from $kl^*_\text{550nm}
\approx 2.1$ up to $kl^*_\text{650nm} \approx 3.45$, thus spanning
a range similar to that of the different samples above. At the
highest and lowest wavelengths, the values of $kl^*$ were
interpolated from the accessible values, which is a good
approximation, since for the investigated region $kl^*$ are found to
scale linearly with $\lambda$ (see supplementary material). The result of such a spectral
measurement of a R700 sample ($L = 0.98\, \text{mm}$ and $m =
377\, \text{mg}$) is shown in Fig. \ref{fig:spectral}. For the wavelengths of
\unit{640}{nm} and \unit{650}{nm}, corresponding to the largest values of $kl^*$, $\sigma^2$
does not saturate, which shows that the mobility edge is approached. This allows a direct characterization of the localization transition with a continuous change of the order parameter.

\begin{figure}
\includegraphics[width=\linewidth]{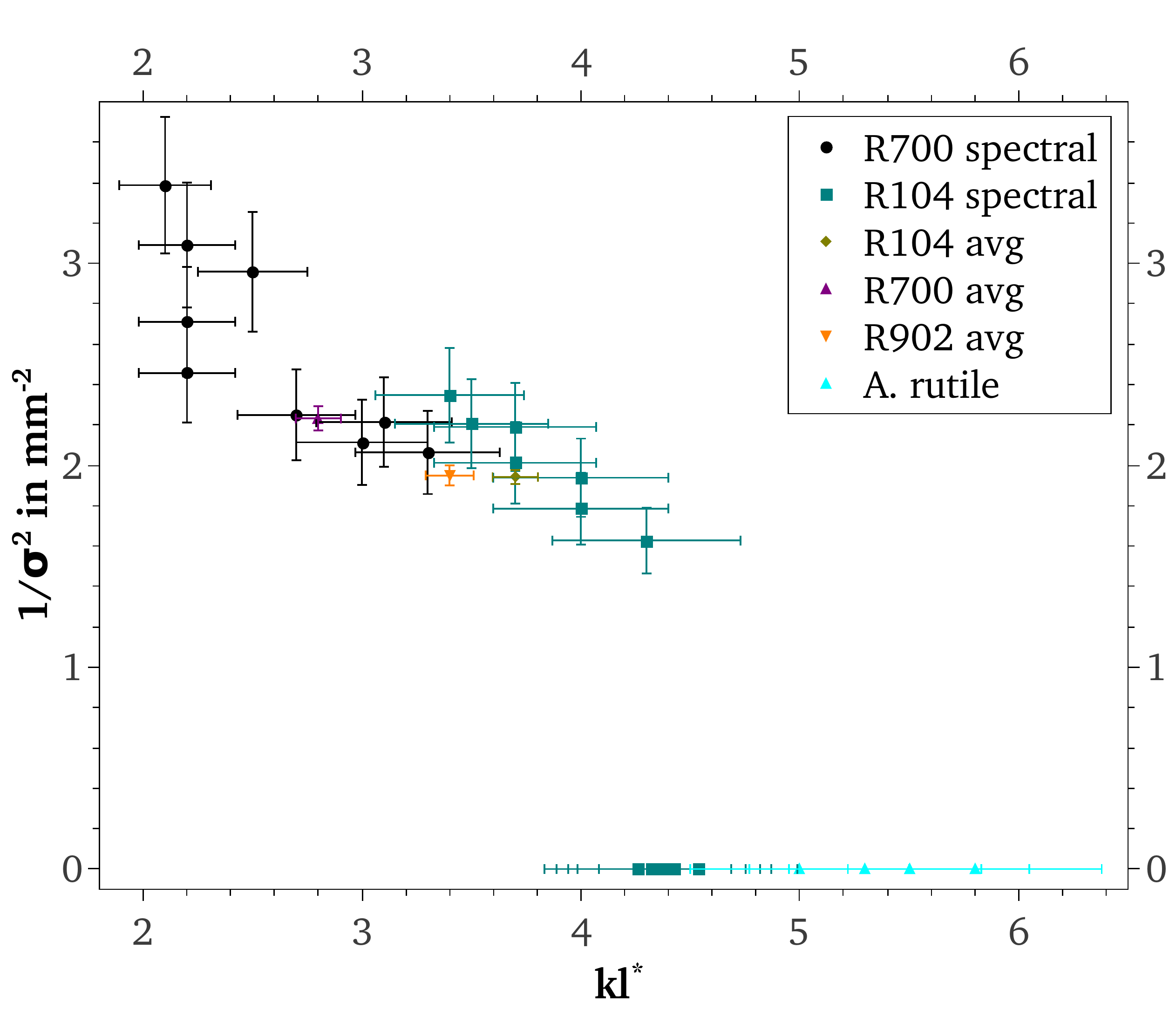}
\caption{\label{fig:kl*-sigma} The inverse of the mean square width $\sigma^2_\infty$ of the plateau against $kl^*$ for different samples. As can be seen, the width, corresponding to the localization length diverges at a value of $kl^* \simeq 4.5$, indicating the transition from a localized to a non-localized state. The increase of the localization length approaching the critical turbidity can also be used to estimate the critical exponent.
}
\end{figure}

\begin{figure}
\includegraphics[width=\linewidth]{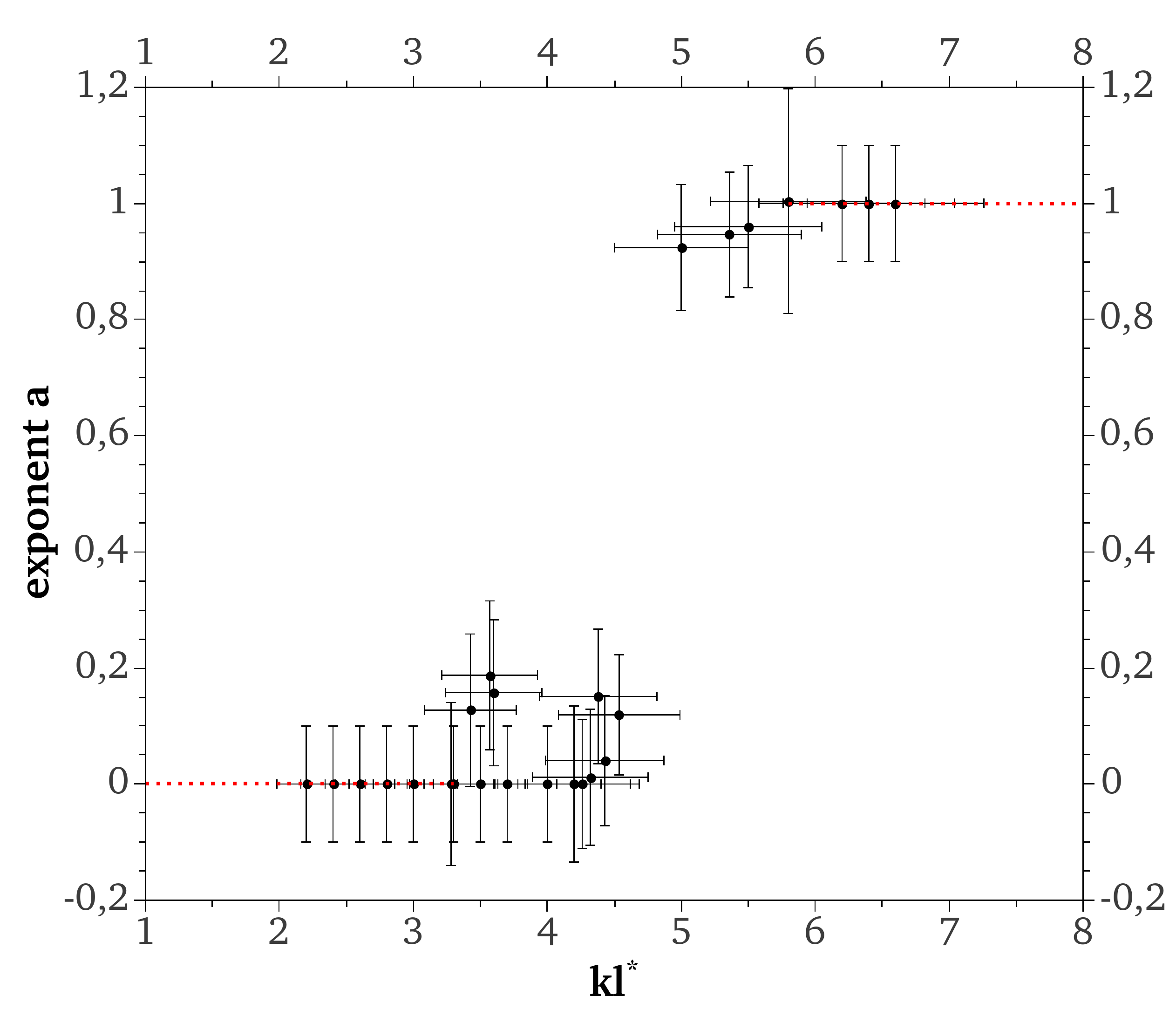}
\caption{\label{fig:transition} The value of the exponent $a$ describing the temporal increase of the mean square width. In the diffusive regime, the exponent should be unity, whereas in the fully localized regime a value of zero is expected. At the mobility edge the sub-diffusive increase corresponds to intermediate values. This allows a determination of the critical turbidity.
}
\end{figure}

We have determined the same spectral information also from a R104
sample, which is closer to the mobility edge at a wavelength of
590 nm and for a rutile sample from Aldrich, which shows classical diffusion at 590 nm. For all of these samples, we have determined the
 value of $kl^*$ \cite{gross07}.
With the value of $\xi$, and the scattering strength
$kl^*$ we are able to determine the approach to the mobility edge
at $kl^*_\text{crit}$, as shown in Fig. \ref{fig:kl*-sigma}. At the mobility edge, we can determine the qualitative change in behavior from the ratio of the slopes of $\sigma^2$ as a function of time in the localized or sub-linear regime and the initial diffusive regime (see supplementary material). This gives a direct estimate of the exponent $a$ with which the width increases with time, $\sigma^2 \propto t^{a}$ shown in Fig. \ref{fig:transition}. There is a clear transition in the behavior with $kl^*$, showing a critical value of $kl^*_\text{crit}=4.5(4)$, above which $a = 1$ and below which $a = 0$. This is in good accord with the determination from time of flight measurements on similar samples yielding
$kl^*_\text{crit,ToF}=4.2(2)$ \cite{stoerzer06n2}. Note that with
an effective refractive index of the samples of $n_\text{eff} \simeq
1.75$, a critical value of $kl^*_\text{crit} = 4.2$ corresponds to
an onset of localization at the point of $l^*/\lambda_\text{eff} = 1$,
which is a reasonable expectation for the onset of localization.

The dependence of the inverse width on the turbidity, as
shown in Fig. \ref{fig:kl*-sigma}, also indicates the critical behavior around the transition. Below the critical turbidity, $\sigma^2$ increases at all times and the corresponding inverse localization length is zero. At the mobility edge, the localization length is limited by the sample thickness, which in the case shown here was approximately 1 mm and a more detailed determination of the intrinsic localization length is not possible. For highly
turbid samples, well below the transition, the inverse localization length
seems to increase linearly with decreasing $kl^*$ indicating an exponent of
unity. However, there is insufficient
dynamic range close to the transition for a full determination of
a critical exponent.

In conclusion, we have shown direct evidence for localization of light in three dimensions and the corresponding transition at the mobility edge. This has been achieved using the time dependence of the mean
square width $\sigma^2$ of the transmission profile, which is an excellent
measure for the onset of localization of light. In contrast to other measures, it is completely independent of absorption and allows a {\em direct} determination of the localization length for samples close to the mobility edge. We find that for highly turbid samples, $\sigma^2$ shows a plateau, which changes to a sub-linear increase for critical turbidities and becomes linear for purely diffusive samples.  This allows a detailed characterization of the behavior of transport close to the transition, which is not possible with other techniques. By evaluating the plateau $\sigma^2_\infty$ of
localizing samples one can directly access the localization length
$\xi$. For sample thicknesses close to the localization length, we moreover observe a decrease in the width of the photon cloud, which we associate with a statistical distribution of microscopic localization lengths. A description of these data will stimulate further theoretical work and comparison between such quantitative theories, such as self-consistent theory \cite{cherroret10} or direct numerical simulation \cite{gentilini10} and the data can then yield valuable information about the statistical distribution of localization lengths close to the transition.

In addition, we have shown that the
transition to localization can be observed in one and the same
sample using spectral measurements, thus continuously varying the control parameter of turbidity through the transition. For highly turbid samples, the width of the
transmission profile saturates at a value, which increases with decreasing
turbidity until the localization length is comparable to the
sample thickness. At this point the width increases at all times,
albeit with a sub-linear increase at long times. This behavior
is expected from the diffusion coefficient at the mobility
edge \cite{berkovits}. Such measurements close to the transition
between Anderson localization and diffusion allow a determination
of the critical turbidity $kl^*_\text{crit} = 4.5(4)$, which is in
good agreement with an indirect determination using time of flight
measurements. In addition, our determination of the localization
length during the approach to the localization transition allows
an estimate of the critical exponent of the transition. Well away
from the critical regime, we find a value close to unity, which is
not incompatible with theoretical determinations
\cite{abrahams,john84,numeric}. A complete description of the transition in open media taking finite size effects into account will be a great challenge for future theoretical descriptions of Anderson localization.

{\bf Methods}

The samples are slabs made up
of nano particles of sizes ranging from 170 to 540 nm in diameter with polydispersities ranging between 25 and 45 $\%$. Powders were provided by DuPont and Sigma Aldrich. These samples are slightly compressed and
have been used previously \cite{stoerzer06} to demonstrate
non-classical transport behaviour in time dependent transmission.
TiO$_2$ has a relatively high refractive index in the visible of $n=2.7$ in
the rutile phase and 2.5 in the anatase phase.

The extremely high turbidity of the samples implies the use of a high power laser system to be able to
measure this transmitted light. We use a
frequency doubled Nd:YAG laser (Verdi V18), operated at
\unit{18}{W} output power, to pump a titanium sapphire laser (HP
Mira). The HP Mira runs mode locked with a repetition rate of
\unit{75}{MHz} at a maximum of about \unit{4}{W}. To convert the
laser light from about \unit{790}{nm} to orange laser light
(\unit{590}{nm}) a frequency doubled OPO is used. The laser wavelength emitted by the OPO can be tuned from approx. \unit{550}{nm} to
\unit{650}{nm}.

To approximate a point-like source the laser beam was focused onto the flat front surface of the sample with a waist of 100 $\mu$m. The
transmitted light was imaged from the flat backside by a magnifying lens ($f
=$\unit{25}{mm}, mounted in reverse position) onto a high rate
intensifier (HRI, LaVision PicoStar). The HRI can be gated on a
time scale of about \unit{1}{ns} and the gate can be shifted in
time steps of \unit{0.25}{ns}. The HRI is made of gallium arsenide
phosphide which has a high quantum efficiency of maximum 40.6 \%
at about \unit{590}{nm}. A fluorescent screen images the signal
onto a \unit{16}{bit} CCD Camera with a resolution of 512 $\times$
512 pixel. With this system we were able to
record the transmitted profile with a time resolution below a
nanosecond.

To measure the turbidity of a sample we used a
backscattering set-up described elsewhere \cite{fiebig08,
gross07}. With this setup covering the full angular range, it is
possible to determine $kl^*$ from the inverse width of
the backscattering cone. Since this system used different laser
sources, the spectral range of the set-up is
more limited in wavelength (\unit{568}{nm} to \unit{619}{nm} and
\unit{660}{nm}).

This work was funded by DFG, SNSF, as well as the Land
Baden-W\"urttemberg, via the Center for Applied Photonics.
Furthermore we like to thank Nicolas Cherroret for his support and
fruitful discussions. 

\newpage

\begin{thebibliography}{99}
\bibitem{HRI} The finite gating window leads to an averaging of the time-dependent width weighted by the time-dependent transmitted intensity. Due to the non-linear time-dependence of this intensity, the averaged width can be different around the maximum intensity for longer gating times.
\end{thebibliography}
\end{document}